\documentclass[aps,prb,twocolumn,groupedaddress,floatfix,showpacs]{revtex4-1}
\usepackage{graphicx}
\usepackage{dcolumn}% Align table columns on decimal point
\usepackage{bm}% bold math
\usepackage{color}% Color!
\usepackage{longtable}% long tables (experimental)

\begin{document}

\title{Low-energy tetrahedral polymorphs of carbon, silicon, and germanium}

\author{Andr\'es Mujica} \affiliation{Departamento de F\'{\i}sica,
  MALTA Consol\'ider Team, Universidad de La Laguna,
  La Laguna 38206, Tenerife, Spain}

\author{Chris J. Pickard} \affiliation{Department of Physics \&
  Astronomy, University College London, Gower Street, London WC1E~6BT,
  United Kingdom}

\author{Richard J. Needs} \affiliation{Theory of Condensed Matter
  Group, Cavendish Laboratory, J J Thomson Avenue, Cambridge CB3 0HE,
  United Kingdom}

\date{\today}

\begin{abstract}
  Searches for low-energy tetrahedral polymorphs of carbon and silicon
  have been performed using density functional theory computations and
  the \textit{ab initio} random structure searching (AIRSS) approach.
  Several of the hypothetical phases obtained in our searches have enthalpies
  that are lower or comparable to those of other polymorphs of group 14 elements
  that have either been experimentally synthesized or recently proposed as the
  structure of unknown phases obtained in experiments, and should thus be
  considered as particularly interesting candidates.
  A structure of $Pbam$ symmetry with 24 atoms in the unit cell was
  found to be a low energy, low-density metastable polymorph in carbon, silicon,
  and germanium.
  In silicon, Pbam is found to have a direct band gap at the zone center with an
  estimated value of 1.4 eV, which suggests applications as a photovoltaic
  material.
  We have also found a low-energy chiral framework structure of
  $P4_12_12$ symmetry with 20 atoms per cell containing fivefold spirals of atoms,
  whose projected topology is that of the so-called Cairo-type two-dimensional 
  pentagonal tiling.
  We suggest that P4$_1$2$_1$2 is a likely candidate for the structure of the 
  unknown phase XIII of silicon.
  We discuss Pbam and P4$_1$2$_1$2 in detail, contrasting their
  energetics and structures with those of other group 14 elements,
  particularly the recently proposed P4$_2$/ncm structure, 
  for which we also provide a detailed interpretation as a network of 
  tilted diamond-like tetrahedra.
\end{abstract}


%71.15.Nc  Total energy and cohesive energy calculations
%62.50.-p  High-pressure effects in solids and liquids
%61.66.-f  Structure of specific crystalline solids
%61.50.-f  Structure of bulk crystals
%61.50.Ah  Theory of crystal structure, crystal symmetry; calculations and modeling 

\maketitle

\section{Introduction}
\label{sec_introduction}

The group 14 elements carbon (C), silicon (Si), and germanium (Ge)
have attracted much interest and have been extensively studied.  
These elements have an $s^2p^2$ valence
electronic configuration, which leads to common chemical features, but
also significant differences.  Pure C is found on the Earth mainly in
the graphite and cubic diamond (cd) forms, which exhibit some of the
strongest bonds known in Nature.  Pure Si and Ge also adopt the
diamond structure under ambient conditions, with ideal tetrahedral
coordination, and they are, of course, semiconductors of great
importance in the electronics industry.
Under applied pressure of around 11 GPa both Si and Ge transform into
the $\beta$-Sn-type structure,\cite{nelmes_1998} which has 4+2
coordination and is metallic. Other stable and well researched phases
of these elements exist at higher
pressures.\cite{nelmes_1998,mujica_2003}

Besides their thermodynamically stable forms, a number of
zero-pressure metastable polymorphs have been observed, with both
higher and lower densities than the corresponding diamond phases.  The
lonsdaleite or hexagonal diamond form of carbon, which has a density
similar to cubic diamond, was first observed in meteorite
craters\cite{Frondel_1967} and later synthesized from graphite under
high pressure and temperature,\cite{bundy_1967_hdC} and also by
chemical vapor deposition and other chemical methods.  Hexagonal
diamond has also been reported in Si and Ge after decompression and
heating from the high-pressure phases or after
indentation.\cite{Wentorf_si_bc8_1963,Domnich_2008,Ruffell_2009} Among
the high-density polymorphs, the cubic bc8 phase of Si and tetragonal
st12 phase of Ge were first obtained about fifty years ago as
zero-pressure tetrahedral metastable forms recovered upon
decompression from the respective high-pressure $\beta$-Sn-type
phases.\cite{Wentorf_si_bc8_1963,bundy_1963_st12Ge,kasper_1964_bc8Si}
The r8 structure (a rhombohedral distortion of the bc8 structure) was
later discovered experimentally in high-pressure experiments on
quenched Si \cite{Piltz_r8_silicon} and has recently been identified
in nanoindented Ge samples.\cite{Johnson_r8_germanium}  Although the
bc8 structure has also been reported in Ge,\cite{Nelmes_1993_Ge-bc8,mujica_1993}
the st12 structure has not so far been observed in Si.\cite{needs_1995}  Low density
open framework clathrate-type structures of Si and Ge, with several
potential applications, can also be produced at ambient pressure using
chemical synthesis methods and have been the object of intense
research in recent years.
\cite{KasperHPC65,San-MiguelKBMPIPRCP99,BeekmanN08,GuloyRTSBG06} There
are also several reports of other Si and Ge phases obtained upon
pressure release in diamond anvil cells or by nanoindentation and
whose structure has not yet been experimentally
resolved.\cite{Zhao_1986_Si-IX,Kailer_1997,Domnich_2002,Ge_2004,Domnich_2008,Ruffell_2009}

Carbon, with the versatility that stems from its unique ability to
form $sp$, $sp^2$, and $sp^3$ hybrid bonds, provides most examples of
allotropes with a wide range of structures and properties.  These
allotropes include, besides those already mentioned, fullerenes,
nanotubes, and graphene and its close
relatives.\cite{Geim_2007_graphene} 
New structural forms have been added to this list in recent times.
A low-temperature quenchable
transparent and superhard metastable crystal phase of carbon, for
which several structures have been proposed, can be obtained by cold
compression of graphite,\cite{mao_2003} whereas the application of
pressure under different controlled conditions to fullerenes and
carbon nanotubes leads to the formation of some new and elusive
forms.\cite{yamanaka_2006,popov_2002,wang_2004}
These findings have excited the imagination and attracted the attention of
researchers to this field and it is expected that new experimental observations
will occur in the coming years.

Along with experimental investigation of observed allotropes, great
effort has been expended in theoretical searches for new tetrahedral
phases whose predicted mechanical, structural and optical properties
may have potential in advanced technological applications. Such
theoretical experiments may result in a fruitful interplay between
theory and practice.  There is a need for computer simulations that
help in characterizing new phases obtained in the laboratory precisely, 
as such phases are difficult to study in high-pressure experiments, 
and simulations can significantly help in identifying the correct 
structure, or in discarding others.

A number of novel hypothetical phases have been proposed from these 
theoretical studies.
One driving force in the search for new carbon phases has been the
quest for superhard materials - novel materials with hardness
rivalling or even exceeding that of carbon diamond, which could be of
technological
importance.\cite{li_2009,Lyakhov_2011_carbon,Zhao_2011_Cco-C8,niu_2012,zhu_2011}
For silicon, the main material used so far in the fabrication of solar
cells, there is interest in new phases with electronic band structures
and optical properties better suited than the diamond phase for
photovoltaic applications.\cite{botti_2012,xiang_2013} There is also a
purely fundamental interest in researching the possibilities of
structural and functional diversity of this important class of
elemental materials.

Here we describe a number of hypothetical energetically favourable
polymorphs of C, Si, and Ge, including two particular low-energy
polymorphs of space group symmetries $Pbam$ and $P4_12_12$.  These
structures were obtained in density-functional-theory (DFT) based
searches.  They are somewhat less dense than the corresponding diamond
structures and have very similar densities to a low-energy P4$_2$/ncm
structure that was recently found in DFT structure searches
\cite{Zhao_2012} and has been considered as a candidate for the
observed metastable phase XIII of
Si.\cite{Kailer_1997,Domnich_2002,Ge_2004,Domnich_2008,Ruffell_2009}
The Pbam structure has a somewhat lower energy in C, Si, and Ge than
the P4$_1$2$_1$2 and P4$_2$/ncm structures.  It may be hoped that some
of these low-energy polymorphs can be synthesized experimentally.

The rest of the paper is organized as follows. In Sec.\
\ref{sec_calculation_details} we give details of our {\it ab initio}
calculations and the structural searches performed. In Sec.\
\ref{sec_results} we show our results for the materials considered,
including detailed descriptions of several selected structures.
Finally, in Sec.\ \ref{sec_summary} we present a summary of our most
important conclusions. The plentiful data generated during the study
has made it necessary to place a portion of them as Supplemental
Material to the present paper.\cite{Supplemental}

\section{{\it Ab initio} Calculations and Structure Searching}
\label{sec_calculation_details}

We performed computational searches for low-energy structures of C and
Si using first-principles density-functional-theory (DFT) methods and
the \textit{ab initio} random structure searching (AIRSS)
approach.\cite{Pickard_silane_2006,Pickard_2011_AIRSS_review} We did
not perform searches for Ge, instead we took the low-energy Si
structures and re-relaxed them for Ge, as low-energy polymorphs of Si
are also expected to be low-energy polymorphs of Ge.  In the AIRSS
approach an ensemble of randomly chosen initial structures are relaxed
to a minima of the energy.  AIRSS has been successfully employed in
finding low-energy structures in many systems, including group 14
elements and their
compounds.\cite{Pickard_alh3_2007,Csanyi_2007_graphite,Morris_2009_defects_in_silicon,Pickard_2010_CFS_structure,Sun_2011_CO,Martinez-Canales_carbon_2012}
We performed searches for structures with up to 24 atoms per unit cell.

The \textsc{castep}\cite{ClarkSPHPRP05} DFT code and the
Perdew-Burke-Ernzerhof (PBE)\cite{PerdewBE96} Generalized Gradient
Approximation (GGA) density functional were employed for the searches.
We used ultrasoft pseudopotentials\cite{Vanderbilt90} with the outermost four
valence electrons treated explicitly   
and default basis set energy cutoffs, further relaxing the
structures of interest at a higher level of accuracy, using a ${\bf k}$-point grid of
spacing $2\pi\times$0.03~\AA$^{-1}$ for the Brillouin zone integrations. 
The details of this procedure are similar to those reported in previous
studies.\cite{Pickard_silane_2006,Pickard_2011_AIRSS_review}

The structures were subsequently re-relaxed using the PBE functional 
and the Vienna {\it ab initio} simulation package
(\textsc{vasp}).\cite{kresse_1993,kresse_1996} For Ge we tested
pseudopotentials including the four valence electrons explicitly and
pseudopotentials in which the $3d$ electrons of Ge were also included
explicitly, but the effects of including the $3d$ electrons were very
small, and we chose to perform phonon calculations with the $d$
electrons treated as core states.  Calculations of the energies as a
function of volume, phonon spectra, and band structures were performed
using the projector augmented-wave (PAW)
method.\cite{Blochl_1994,kresse_1999} The phonon calculations were
performed with the \textsc{phon} code \cite{Alfe_phon}.  The plane
wave cutoff energies used for the \textsc{vasp} calculations were 520
eV (C), 320 eV (Si), and 230 eV (Ge, 375 eV for the calculations with
the $3d$ electrons). Dense ${\bf k}$-point grids were used for the Brillouin
zone integrations (e.g., an 8$\times$8$\times$8 grid was used for the
Pbam structure, with 24 atoms per cell).  The structural relaxations
were deemed to be converged when all of the forces were less than 5
meV \AA$^{-1}$ and the anisotropy of the stress tensor was less than
0.1 GPa.  The results of the \textsc{castep} and \textsc{vasp}
calculations were in excellent agreement with one another, which is a
useful crosscheck.  We also performed \textsc{vasp} calculations using
the local spin-density approximation (LSDA)\cite{ceperley_1980} and
PBEsol density functionals,\cite{PBEsol} which gave qualitatively very
similar results to the PBE calculations and allowed us a full
comparison among the three functionals.

We calculated the total energy $E$ as a function of volume $V$ for
each phase and fitted the data using a fourth-order Birch-Murnaghan
equation of state\cite{birch_1947} (EoS) from which the pressure, $p$,
and the enthalpy, $H=E+pV$, were obtained.

\section{Results}
\label{sec_results}

\subsection{Energetics}
\label{results_energetics}

\begin{figure}[h!]
  \includegraphics[width=0.45\textwidth]{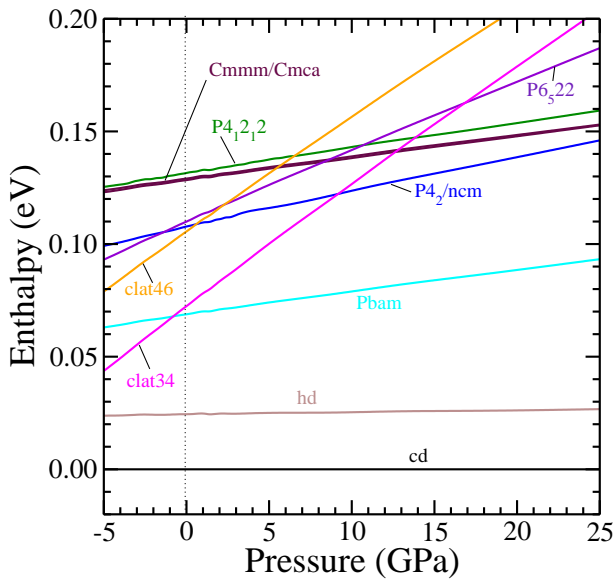}
  \caption{\label{fig:enthalpy-pressure_carbon} (Color online)
    Enthalpy-pressure relations for various structural phases of carbon calculated
    using the PBE.  The enthalpies are given with respect to the
    zero-pressure C-cd phase.}
\end{figure}

\begin{figure}[h!]
  \includegraphics[width=0.45\textwidth]{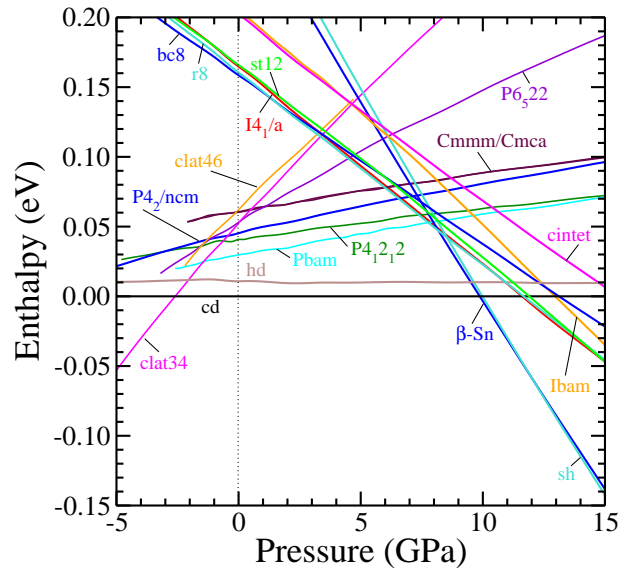}
  \caption{\label{fig:enthalpy-pressure_silicon} (Color online)
    Enthalpy-pressure relations for various structural phases of silicon calculated
    using the PBE.  The enthalpies are given with respect to the
    zero-pressure Si-cd phase.}
\end{figure}

\begin{figure}[h!]
  \includegraphics[width=0.45\textwidth]{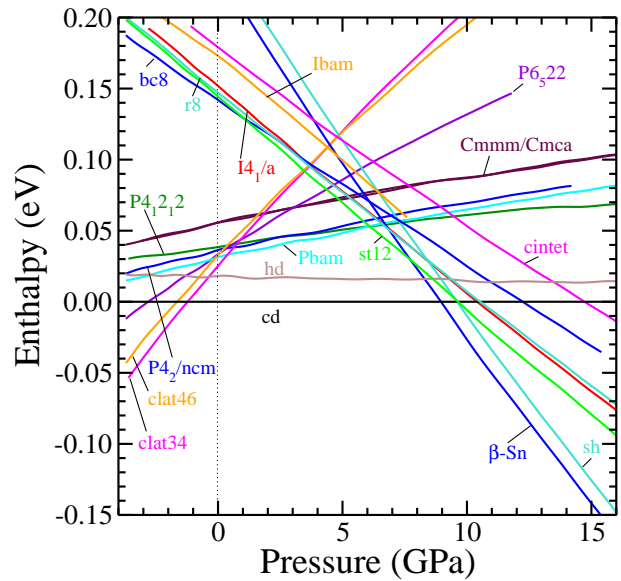}
  \caption{\label{fig:enthalpy-pressure_germanium} (Color online)
    Enthalpy-pressure relations for various structural phases of germanium calculated
    using the PBE.  The enthalpies are given with respect to the
    zero-pressure Ge-cd phase.}
\end{figure}

Enthalpy-pressure relations 
for a selection of different structures in C, Si, and
Ge are shown in Figs.\ \ref{fig:enthalpy-pressure_carbon},
\ref{fig:enthalpy-pressure_silicon}, and
\ref{fig:enthalpy-pressure_germanium}.  Calculated equilibrium
zero-pressure volumes $V_0$, bulk moduli $B_0$, and pressure
derivatives of the bulk modulus $B_0^{\prime}$, obtained from the EoS
fitting, are collected in Tables \ref{table:properties_carbon},
\ref{table:properties_silicon}, and \ref{table:properties_germanium},
for C, Si, and Ge, respectively.  The differences in energy from the
corresponding diamond structures at zero pressure are also reported in
Tables \ref{table:properties_carbon}-\ref{table:properties_germanium}.
For a given structure, these energy differences generally decrease
with increasing atomic number. (See also Supplemental
Material.\cite{Supplemental})

In general, we find good agreement between the calculated and
experimental values of structural parameters for the observed phases.
For example, our values of the volume, bulk modulus and pressure
derivative of the bulk modulus of C-cd are in good agreement with the
experimental values of $V_0 = 5.6738(13)$ \AA$^3$ (room
temperature),\cite{Reeber_diamond_lattice_constant} $B_0 = 444(4)$ GPa
\cite{Occelli_diamond_2003} and $B_0^{\prime} = 3.65(5)$, data from
similar DFT calculations,\cite{Tse_2008_diamond,Kunc_diamond_2003} and
values from accurate quantum Monte Carlo
calculations.\cite{MaezonoMTN07,Casino_reference} We have adopted the
experimental data of Ref.\ \onlinecite{Occelli_diamond_2003} for $B_0$
and $B_0^{\prime}$ of C-cd, with the revised pressure scale of Ref.\
\onlinecite{Tse_2008_diamond}.  Likewise, we obtain good agreement
with the experimental data for Si-cd and Ge-cd.

\subsection{Overview of the findings}
\label{searches_results}

It is normally possible to identify families of related crystalline
structures.  For example, the cd and hd structures are members of an
infinite family of low-energy polytypic structures consisting of
different stackings of layers.  Structures consisting of (1) different
stackings of layers; (2) regions of the most stable cd phase and
interfacial regions; (3) periodic arrays of point defects in the cd
structure etc., can be dreamt up with energies per atom arbitrarily
close to that of the corresponding cd structure.  It is not our
intention to investigate such families of tetrahedrally-bonded group
14 structures, but instead we seek structures which differ in
substantial ways from known structures.

Our searches produced many structures, including the observed
equilibrium cubic diamond (cd) structure of C and Si, the hexagonal
diamond structure (hd),\cite{footnote_7} and dense structures adopted
by Si under high pressures, such as the $\beta$-Sn structure.  As well
as finding experimentally observed tetrahedrally-bonded phases we
found many structures that have been reported in recent DFT searches
of group 14 elements 
and considered others explicitly for the sake of comparison. 
\cite{Pickard_2010_CFS_structure,Lyakhov_2011_carbon,Malone_2012_Si_Ibam,Malone_2012_Ge,Zhao_2012,Selli_germanium}
(Though we will not report on all of them, some further data can be
found in the Supplemental Material.\cite{Supplemental})
We refer to the new structures by the names of their space groups 
and use common abbreviations for the well
established structures, 
although some structures will be occasionally also referred to by other names with which they
have appeared in the recent literature on this subject.
We focus here on tetrahedrally bonded low-energy structures.

\subsubsection{High-density tetrahedral polymorphs}

In our searches we found all the dense tetrahedral polymorphs that
have been observed in Si and/or Ge: the bc8 and r8 structures obtained
upon decompression from the corresponding high-pressure $\beta$-Sn
phase or nanoindentation,
\cite{Wentorf_si_bc8_1963,Nelmes_1993_Ge-bc8,Piltz_r8_silicon,Johnson_r8_germanium,mujica_2001}
and the st12 structure which has only been observed upon decompression
in Ge.\cite{bundy_1963_st12Ge,mujica_1993,needs_1995} We also found a body-centred tetragonal
structure with space group $I4_1/a$ and 8 atoms per primitive cell,
whose enthalpy is remarkably close to that of r8/bc8, and which
plays a role in the decompression kinetics from the high pressure phases.\cite{wang_2013}
%, to the best of our knowledge, has not been reported before.  
For Si
within the PBE, for example, I4$_1$/a is only 5 meV per atom higher in
enthalpy than bc8/r8 at zero pressure (and about the same enthalpy as
st12, within the accuracy of our calculations), see Table
\ref{table:properties_silicon}, which makes I4$_1$/a an energetically
competitive high-density polymorph.

Also belonging with the group of high-density tetrahedral polymorphs,
an hexagonal structure of $P6_422$ symmetry labelled cintet is
interesting as it is the elemental analog of the tetrahedral variant
of the binary cinnabar structure (also called pseudo-cinnabar structure)
observed at high pressures in both ZnTe and
GaAs.\cite{nelmes_1998,mujica_2003,mujica_1998} At zero pressure, the cintet
structure in Si and Ge is very close in energy to the Ibam
structure with 6 atoms per primitive cell recently proposed by Malone
and Cohen.\cite{Malone_2012_Si_Ibam} They are both, however, higher in
energy than the other observed dense allotropes bc8/r8/st12, which may
impede their direct synthesis as metastable phases on pressure release
from the high-pressure $\beta$-Sn phase.  At higher pressures Ibam is
lower in enthalpy than cintet.  In general, the high-density
allotropes have equilibrium energies significantly higher than those
of the low-density forms that we discuss next.  However, their reduced
volumes make them competitive at high pressures, and indeed it is
under such conditions that some of them have been synthesised.
Among the phases considered, the cintet phase has a bulk modulus second only to
the diamond forms of C, and somewhat higher to their respective diamond forms
in Si and Ge.

\subsubsection{Low-energy and lower-density tetrahedral polymorphs}

One interesting result of our search is that we have found a C-centred
orthorhombic structure of $Cmca$ symmetry containing 16 atoms in the primitive
unit cell whose $E(V)$ curve is almost degenerate, in each of the group 14
elements studied, with that of the Cmmm structure (with 8 atoms per primitive
cell) previously found by Zhao {\it et al.}\cite{Zhao_2011_Cco-C8} in carbon
and named Cco-C$_8$ by these authors.  This Cmmm structure has been proposed as
that of a superhard carbon allotrope recovered from room-temperature
compression of carbon nanotube (CNT) bundles,\cite{wang_2004} and it is indeed
structurally related to the (2,2) CNT.\cite{Zhao_2011_Cco-C8} The Cmmm carbon
phase of Zhao {\it et al.}\ was also readily found with AIRSS.

From our calculations, both the Cmmm and Cmca structures are very
close in energy to the Pmmn structure of the so-called $P$ carbon
phase, with 16 atoms per cell, alternatively proposed by Niu {\it et
  al.}\cite{niu_2012} as the structure of the cold-compressed CNT
obtained in experiment in Ref.\ \onlinecite{wang_2004}.  Both the new
Cmca allotrope and the Cmmm and Pmmn allotropes share similar
structural motifs, namely variations on arrangements of the same
four-, six-, and eightfold ring patterns, which is the origin of their
similar energetics. They also contain many sixfold diamond-like rings
which may partially account for their relatively low energy and
superhardness in carbon. Possibly, other energetically close
structures could be conceived exhibiting variations of these
motifs. This suggests that cold compressed CNT bundles could adopt a
mix of different but structurally and energetically very close forms,
as described by the family Cmmm/Pmmn/Cmca.

Our search also produced the primitive tetragonal structure with 12
atoms per cell (T12) and space group $P4_2/ncm$ that has been recently
reported by Zhao {\it et al.}\cite{Zhao_2012} In the present work we
provide a detailed description of this structure and its relation to
diamond (see Sec.\ \ref{P42ncm_struc}), interpreting P4$_2$/ncm as a
network of diamond-like tetrahedra made up of slabs of almost regular
corner-sharing tetrahedra alternatively rotated left and right, so
that fivefold connectivity appears within each slab and adjacent slabs
are bonded together by non-shared corners.  We believe that this
remarkable structure is the simplest that can be built by the stacking
of slabs of tilted tetrahedra while preserving to a large degree a
highly regular fourfold coordination for the sites.  The P4$_2$/ncm
structure has been proposed\cite{Zhao_2012} as a candidate for both the 
Si-XIII phase (which is observed to coexist along with the
Si-cd, Si-hd, Si-r8, and Si-bc8 phases
% upon release of pressure from the high-pressure metallic
% $\beta$-Sn-type phase and
in nanoindentation
experiments\cite{Kailer_1997,Domnich_2002,Ge_2004,Domnich_2008,Ruffell_2009})
and for an experimentally synthesized metastable Ge
phase,\cite{Saito_1978} and for this reason we compare our data for
other candidate phases with those of P4$_2$/ncm.\cite{footnote_9}

We have also obtained two interesting and novel low-energy structures,
Pbam (with 24 atoms per cell) and P4$_1$2$_1$2 (20 atoms per cell),
with orthorhombic and tetragonal symmetry, respectively, which are
energetically competitive polymorphs and whose detailed description
will be given below.  The orthorhombic Pbam structure and the
tetragonal P4$_1$2$_1$2 
%and P4$_2$/ncm structures 
structure are slightly
expanded in volume with respect to the corresponding diamond
structures, with zero-pressure equilibrium volumes 
%between about 2 and 5 \% larger 
about 2-3 \% larger
than those of the corresponding cd structures, see Tables
\ref{table:properties_carbon}-\ref{table:properties_germanium}.  The
density of the P6$_5$22 chiral framework structure reported previously
\cite{Pickard_2010_CFS_structure} is intermediate between those of
Pbam/P4$_1$2$_1$2/P4$_2$/ncm and the type-I and II clathrates.  The calculated
phonon dispersion curves of these phases show that they are
dynamically stable.

Note that the present Pbam structure with 24 atoms per cell is not the same as
the Pbam structure with 16 atoms per cell proposed by Niu {\it et al.} for $R$
carbon,\cite{niu_2012} along with other orthorhombic structures.  It is in
particular significantly lower in energy.  The present Pbam structure is
energetically highly favorable in C, Si, and Ge.  For carbon, Pbam is
essentially degenerate in energy with the most favourable of the low-symmetry
defect structures proposed recently by Botti {\it et al.}, with monoclinic
symmetry and 20, 22, and 24 atoms per cell,\cite{botti_2013} whereas in our
study for silicon and germanium we have considered \textit{ex professo} these
hypothetical carbon structures and have found them to lie within 4 meV of Pbam,
which is close to the accuracy of the calculations.\cite{footnote_6} Apart from
these structures, among those considered here, only the diamond forms and
the type-II clathrate of germanium (Ge-clat34) are significantly lower in 
%energy
enthalpy at zero pressure
than the corresponding Pbam structures, with the type-I germanium clathrate
(Ge-clat46) having about the same 
%energy
zero-pressure entahlpy as Ge-Pbam (and a rather smaller
density).

Low density clathrate structures of types I and II 
(here labelled clat46 and clat34, respectively) 
have been synthesized in Si, Ge, and Sn,
\cite{KasperHPC65,San-MiguelKBMPIPRCP99,BeekmanN08} although not in
C.  Group 14 clathrates have normally been synthesized by
incorporating guest atoms, although a guest-free Ge clathrate has been
formed.\cite{GuloyRTSBG06}  Our calculated results for the clathrates
are in good agreement with earlier work.
\cite{ReyMHM08,DongS99,MylesBS01,Karttunen_2011,Adams_1994_clathrates}  The high-density
polymorphs bc8, r8 and st12 of Si and Ge are all considerably less
stable than Pbam at zero pressure.

The P4$_1$2$_1$2 form is a chiral framework structure containing
fivefold spirals of atoms, that is also energetically competitive,
particularly in Si, for which among the new polymorphs only the Pbam structure
is lower in energy.  In C, its enthalpy is very close to those of the
Cmmm/Cmca phases already mentioned in connection with cold compressed
nanotubes. The P4$_2$/ncm structure is less stable than P4$_1$2$_1$2
in Si, but more stable than P4$_1$2$_1$2 in C 
whereas for Ge both structures have approximately the same enthalpy.
%and only marginally more stable in Ge. 
The two pure
clathrate structures considered are also less stable than P4$_1$2$_1$2
in Si.

As has been observed previously, the calculated energy differences
within the PBE are systematically larger than those from the PBEsol
functional, and these are larger in turn than those from the LDA.  One
effect of this is that the coexistence pressures with the diamond
phases for the high pressure phases (such as $\beta$-Sn) or the
high-density tetrahedral metastable phases (such as bc8) are larger
with PBE than PBEsol or the LDA.  The differences in energy between
the PBE and the LDA or PBEsol are, however, quite small for the
low-density tetrahedral polymorphs with equilibrium volumes close to
the diamond phase such as Pbam, P4$_2$/ncm, or P4$_1$2$_1$2, whereas
for the high-density tetrahedral polymorphs such as bc8, r8, or st12
there is a significant difference between the results obtained with
different exchange-correlation approximations, see Tables
\ref{table:properties_carbon}-\ref{table:properties_germanium} and
Supplemental Material.  This reflects the similarity in bonding in
diamond and in phases such as Pbam, P4$_2$/ncm, or P4$_1$2$_1$2.  
The bulk moduli of these and the other low-density structures obtained in the
searches tend to be slightly below the bulk moduli of the corresponding diamond
forms. In the case of carbon, for example, these are in the range $\sim$400-415 GPa
(cf. the calculated value for C-cd of 435 GPa).

%NOTE bc8 etc not shown in this figure, as they are very high in energy
%NOTE only a few phases shown

\subsection{Description of selected low-energy structures}
\label{structures}

The three phases on which we focus next, P4$_2$/ncm, Pbam, and P4$_1$2$_1$2,
are structurally very different among themselves and with respect to the
diamond structure.  Topologically, all the atoms in these structures have a
tetrahedral coordination of nearest neighbours, in some cases considerably
distorted from a regular coordination, presenting five-, six-, and/or sevenfold
rings of atoms.  Each of these novel structures has similar characteristics in
all four materials studied.  Hereafter we will discuss them in detail
providing, for the sake of illustration, calculated numerical data (values of
distances, angles etc) which correspond to Si at zero pressure, unless
otherwise stated.  In Table \ref{table_crystalstructures} we summarize the
relevant crystallographic structural data for the P4$_2$/ncm, Pbam, and
P4$_1$2$_1$2 structures of Si, at zero pressure, calculated using the PBE
functional. 
%\cite{footnote_5}
Diffraction patterns of the structures simulated using the \textsc{fullprof}
software\cite{FULLPROF_ref} are shown in Fig.\ \ref{fig:XRD_patterns}.
Further details of these structures are given in the Supplemental
Material.\cite{Supplemental}

\begin{table}
\centering
\caption{
  Crystallographic data for the I4$_1$/a, Cmca, P4$_2$/ncm, Pbam, 
  and P4$_1$2$_1$2 structures of Si at zero pressure, calculated 
  using the PBE: space group (SG), lattice parameters, and
  atomic sites (Wyckoff positions).  Those internal coordinates 
  fixed by symmetry are given as integer fractions, to aid recognition.  
}
\begin{ruledtabular}
\begin{tabular}{llll}
\\
I4$_1$/a   & \multicolumn{3}{l}{SG $I4_1/a$, No.88}
\\
           & \multicolumn{3}{l}{$a$=6.676 \AA , $c$=6.514 \AA }
\\
 & Si1 & 16f & (-0.0962, -0.0994, 0.1170)  \\
\\
\hline
\\
Cmca       & \multicolumn{3}{l}{SG $Cmca$, No.64}
\\
           & \multicolumn{3}{l}{$a$=6.480 \AA , $b$=15.496 \AA , $c$=6.693 \AA }
\\
 & Si1 & 16g & (0.3161, 0.0673,  0.4139)    \\
 & Si2 & 16g & (0.3154, 0.1887, -0.0849)    \\
\\
\hline
\\
P4$_2$/ncm & \multicolumn{3}{l}{SG $P4_2/ncm$, No.138}
\\
           & \multicolumn{3}{l}{$a$=5.221 \AA , $c$=9.295 \AA }
\\
 & Si1 & 4b & (3/4, 1/4, 1/4)                   \\
 & Si2 & 8i & (0.0865, 0.0865, 0.3924)          \\
\\
\hline
\\
Pbam       & \multicolumn{3}{l}{SG $Pbam$, No. 55}
\\
           & \multicolumn{3}{l}{$a$=11.774 \AA , $b$=11.081 \AA , $c$=3.863 \AA }
\\
 & Si1 & 4h & (0.6158, 0.2400, 1/2)  \\
 & Si2 & 4h & (0.4625, 0.1005, 1/2)  \\
 & Si3 & 4h & (0.9081, 0.9569, 1/2)  \\
 & Si4 & 4g & (0.7705, 0.7075, 0)  \\
 & Si5 & 4g & (0.8040, 0.0124, 0)  \\
 & Si6 & 4g & (0.1501, 0.6240, 0)  \\
\\
\hline
\\
P4$_1$2$_1$2 & \multicolumn{3}{l}{SG $P4_12_12$, No. 92}
\\
             & \multicolumn{3}{l}{$a$=8.903 \AA , $c$=5.261 \AA }
\\
 & Si1 & 8b & (0.4312, 0.1390, 0.7001) \\
 & Si2 & 8b & (0.4752, 0.6504, 0.4205) \\
 & Si3 & 4a & (0.2011, 0.2011, 1/2) \\
\\
\end{tabular}
\end{ruledtabular}
\label{table_crystalstructures}
\end{table}

\begin{figure}
 \includegraphics[width=0.4\textwidth]{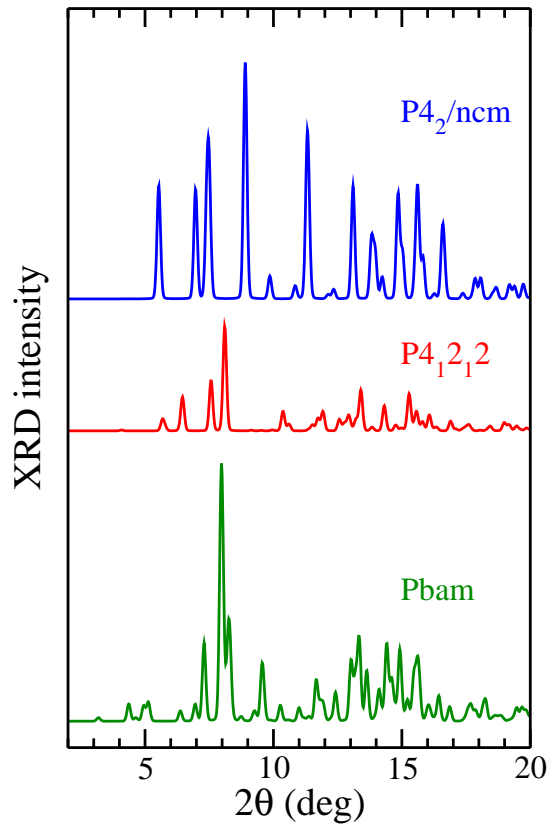}
\caption{\label{fig:XRD_patterns} (Color online) Simulated powder XRD
patterns of the P4$_2$/ncm, P4$_1$2$_1$2, and Pbam phases of Si at
zero pressure, calculated within the PBE approximation. The
x-ray wavelength used was 0.44864 \AA. 
%the same wavelength used by Hanfland in the study of Si-VI
}
\end{figure}

\subsubsection{The P4$_2$/ncm structure as a network of tilted tetrahedra \label{P42ncm_struc}}

\begin{figure}
 \includegraphics[width=0.45\textwidth]{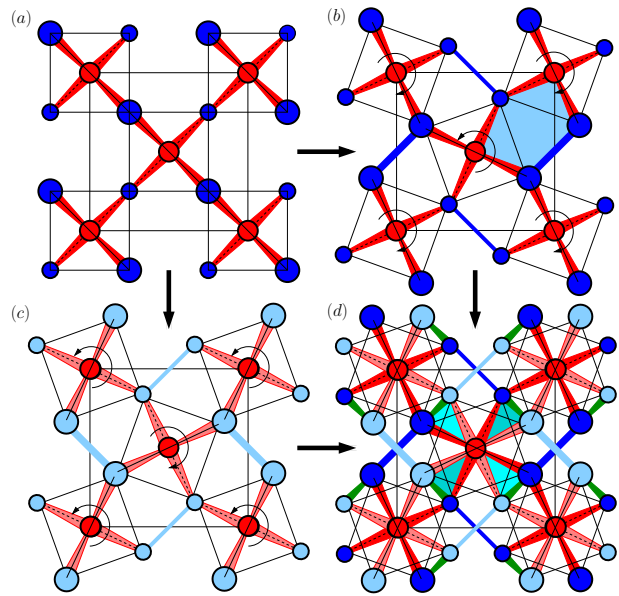}
\caption{\label{fig:structures_P4_2/ncm} Construction of P4$_2$/ncm as
  a network of tilted diamond-like tetrahedra.  Panel (a) shows a
  single slab of corner-sharing regular tetrahedra of the diamond
  structure viewed along the (001) direction.  The projected
  tetrahedra and conventional cubic cell are shown in black.  The
  centers of the tetrahedra are shown in red and their corners in
  blue, with the different sizes of the atoms and varying thickness of
  the bonds used to give a sense of perspective.  Panels (b) and (c)
  show two adjacent slabs ({\it lower} and {\it upper}, respectively)
  of tilted tetrahedra that alternate stacked along the $z$ axis to
  give the P4$_2$/ncm structure, as shown in (d).\cite{footnote_3}
  Note the opposite orientation of the tetrahedral bonds and the
  tilting of the tetrahedra in the two adjacent slabs.  Intra-slab
  bonds (parallel to the plane of projection), resulting from the
  tilting of the tetrahedral units within each slab, are represented
  by blue rods, while inter-slab bonds, arising from the stacking of
  the slabs, are represented by green rods.  Intra-slab bonding leads
  to characteristic fivefold rings, one of which is depicted in blue
  in (b), while inter-slab bonding leads to buckled sixfold rings, two
  of which are shown in profile in (d).  }
\end{figure}

P4$_2$/ncm is a remarkably simple structure that can be understood as
a stacking of (001) slabs of diamond-like corner-sharing tetrahedra,
with adjacent tetrahedra within each slab alternately tilted clockwise
and anti-clockwise by approximately 20$^\circ$, see Fig.\
\ref{fig:structures_P4_2/ncm}.  The centers of the tetrahedra
correspond to the Si1 sites of the P4$_2$/ncm structure (red sites in
Fig.\ \ref{fig:structures_P4_2/ncm}) and their corners correspond to
the Si2 sites (blue).  There are only three free internal parameters
in P4$_2$/ncm, all them related to the Si2 corner sites, whereas the
Si1 center sites form a simple tetragonal lattice.  Within each of
these tetrahedral slabs, the rotation of adjacent tetrahedra in
opposite sense brings their corners together, with the formation of
new bonds between them (blue rods in Fig.\
\ref{fig:structures_P4_2/ncm}b), above and below the plane of the
centers and orientated along $\langle$110$\rangle$ directions.  When
two of these tetrahedral slabs, oppositely orientated (Fig.\
\ref{fig:structures_P4_2/ncm}b and c), are stacked along the $z$ axis,
new inter-slab bonds are established between the corners of each
tetrahedron and the corners of the opposite tetrahedra immediately
above and below it (green rods in Fig.\
\ref{fig:structures_P4_2/ncm}d).  Note that the centers of the
tetrahedra in different slabs are strictly aligned along the stacking
direction, whereas the orientation of the stacked tetrahedra are
alternately tilted clockwise and anti-clockwise.  As a result of
intra- and inter-slab bonding of tetrahedral corners, each Si2 corner
site is linked to two other Si2 corner sites (as well as to two Si1
center sites), with the formation of staggered chains of Si2 corner
sites running along $\langle$110$\rangle$ directions.

It is quite remarkable that while the tilting and rebonding reduces
the symmetry of the P4$_2$/ncm tetrahedral arrangement with respect
diamond, so that in P4$_2$/ncm the tetrahedral units become tetragonal
disphenoids, these units are actually very close to regular
tetrahedra. The Si1 center sites have an almost ideal tetrahedral Si2
environment, with all four bonds of equal length and near-ideal
tetrahedral angles.  The tetrahedral units perform largely as rigid
units in the P4$_2$/ncm structure.

The tilting and subsequent intra-slab corner-bonding leads to the
appearance of fivefold rings made up of two Si1 tetrahedra centers and
three Si2 corners, see Fig.\ \ref{fig:structures_P4_2/ncm}b.  Each of
these fivefold rings is a slightly buckled and symmetrical pentagon
with four out of its five sides having the same length, and with three
different angles, which deviate only moderately from the angle of
108$^\circ$ of a regular pentagon. The stacking of opposite
tetrahedral slabs and accompanying corner-bonding further results in
twisted sixfold rings (2$\times$Si1, 4$\times$Si2) with alternating
orientation along the $z$ axis, noted by Zhao {\it et
  al.},\cite{Zhao_2012} and of sevenfold rings (2$\times$Si1,
5$\times$Si2) in a chair-like configuration, also alternating in
orientation along the stacking direction.  The twisted sixfold rings
are quite unlike those in the diamond structure.  There is only one
type of five-, six-, and sevenfold ring.

There are only three different bond lengths in P4$_2$/ncm: one ($d_1$)
between Si1 and Si2 sites (that is, from the center to each corner of
the quasi-regular tetrahedra), another between neighboring Si2 sites
in the same slab ($d_2$), and a third between Si2 sites in adjacent
slabs ($d_3$).  These three different bond lengths are represented,
respectively, by red, blue, and green rods in Fig.\
\ref{fig:structures_P4_2/ncm}. The {\it tetrahedral} distance $d_1$ is
very similar to that of the diamond phase (2.37 \AA) and shows a
similar compressibility under applied pressure, which is in line with
the fact that the tetrahedral units of P4$_2$/ncm remain very much
diamond-like in nature. The {\it inter-slab} distance $d_3$ is also
similar though it has a slightly larger compressibility, while the
{\it intra-slab} distance $d_2$ is larger than both $d_1$ and $d_3$
and it also shows a larger compressibility.  There are also six
different bond angles,\cite{footnote_1} with only one angle deviating
significantly (122.5$^\circ$) from the ideal tetrahedral value
(109.47$^\circ$).  This large angle corresponds to the chains of bonds
between Si2 tetrahedra corners running along $\langle$110$\rangle$
directions.  The bond angles in P4$_2$/ncm vary very little when the
structure is compressed. The large Si2-Si2-Si2 angle increases by
about 1.5$^\circ$ in the range from -5 GPa to 10 GPa, while the rest
of the angles change by tenths of a degree.  The tilting of the
tetrahedra is locked by intra- and inter-slab bonding and the aspect
of the P4$_2$/ncm structure would not change much even over a
substantial pressure range.

\subsubsection{The Pbam structure}

\begin{figure}
 \includegraphics[width=0.4\textwidth]{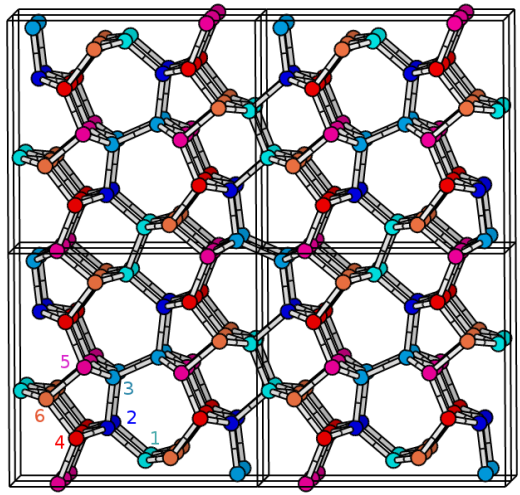}
\caption{\label{fig:structures_Pbam} (Color online) A slab of the Pbam
  structure viewed along the $z$ axis, with unit cells shown in black
  and a modicum of perspective so as to reveal the connectivity among
  sites.  The six inequivalent sites are labelled and shown in
  different colors: the 4g-type sites are depicted in three different
  hues of red, while blue is used for the three 4h-type sites. The
  fivefold and sevenfold rings of the structure (as well as sixfold
  diamond-like rings) are all visible in this projection.  Note that
  the red 4g sites and the blue 4h sites are each in different
  projection planes a distance $c/2$ apart.}
\end{figure}

With six inequivalent sites and twelve free internal parameters, Pbam
has a larger variation of tetrahedral environments than P4$_2$/ncm,
and yet many elements of Pbam remain remarkably similar to those of
diamond with which a useful comparison can be established.  In Fig.\
\ref{fig:structures_Pbam} we represent the Pbam structure in a way
that allows comparison to the diamond structure, when the latter is
viewed along its $\langle$110$\rangle$ direction.  In analogy to
diamond, we see stacks of zigzag chains in Pbam, two per cell, linking
the three 4g-type sites Si4-6 (red sites in Fig.\
\ref{fig:structures_Pbam}).  These distorted diamond-like chains are
contained in (001) projection planes and run along the
$\langle$010$\rangle$ direction.  The three 4h-type sites Si1-3 (blue
sites) form in turn two staggered planar chains along the same
direction but at different (001) planes from the 4g-chains and
intercalated between them.  Also note the relative displacements and
different orientations of consecutive 4g and 4h chains.  Adjacent 4g
and 4h chains are connected by three different types of bond which
form zigzag diamond-like chains running perpendicular to the
projection plane, each linking one type of 4g site to one type of 4h
site (Si1-Si6, Si2-Si4, and Si3-Si5), so each 4g site (respectively,
4h site) is linked to two 4g (4h) sites on the same (001) plane and to
two 4h (4g) sites of the same type, above and below the plane.  The
orientations of these three different connecting 4g-4h chains of bonds
deviate greatly from the regular orientation found in the diamond
structure (which would correspond to a (010) plane in Pbam) but the
chains themselves are very little distorted compared with equivalent
chains in diamond.

The main topological difference between diamond and Pbam arises from
the intra-chain connectivity within the 4h slabs in which drastic
changes lead to the formation of the fivefold and sevenfold rings
shown in Fig.\ \ref{fig:structures_Pbam}.  There are in fact no other
major topological changes either within the 4g slabs or within the
interconnecting 4g-4h chains.  A simple interpretation of Pbam is that
it consists in a certain pattern of shearing of atomic slabs in
diamond such that breaking and formation of intra-slab bonds occurs
{\it only} within alternate slabs (the stacks of 4h chains in Pbam)
whereas the connecting chains of inter-slab bonds are not affected
beyond a small deformation of bond lengths and angles.  All in all,
two bonds are broken and reformed per 4h chain and cell, which
drastically transforms the 4h intra-slab connectivity giving rise to
fivefold and sevenfold rings perpendicular to the shearing direction,
though some sixfold diamond-like rings are also preserved, see Fig.\
\ref{fig:structures_Pbam}.  Whereas the 4h slabs undergo an important
reconstruction, the topology of the 4g slabs remains essentially
unchanged (though deformed) and the 4g-4h chains of connecting bonds
are mostly unaltered (though rotated from their initial orientation by
the intra-4h rebonding).  These changes are accompanied by a
contraction in the $y$ direction and elongation in the $x$ direction
with respect to diamond.

There is one type each of fivefold, sixfold, and sevenfold rings in
the $xy$ projection of Pbam shown in Fig.\ \ref{fig:structures_Pbam},
which appear in four different orientations.\cite{footnote_4} Apart
from the sixfold rings shown in Fig.\ \ref{fig:structures_Pbam}, the
Pbam structure preserves a large number of other sixfold chair-like
rings of the diamond structure perpendicular to the $xy$ plane which
are shown in profile in this figure.  (There are also eightfold rings
perpendicular to the $xy$ plane formed by the rebonding of the 4h
slabs).

Due to the large number of free parameters of the Pbam structure,
there are ten different nearest neighbor distances and twenty one
different bond angles.  Distances vary between 2.33 and 2.42 \AA\
(cf.\ the calculated value for the diamond structure of 2.37 \AA) with
most distances within 0.4 \% of the diamond value and a mean distance
of 2.37 \AA.  Bonds within the 4g slabs are somewhat shorter than
diamond bonds whereas those within the 4h slabs are somewhat longer,
with the longest bond occurring between the Si2 and Si3 sites that
form the pentagonal rings.  Bond compressibility is also similar to
that in diamond, although the two Si1-Si2 and Si3-Si3 bonds within the
4h slabs have somewhat larger compressibility and one Si5-Si6 bond
within the 4g slabs has a smaller compressibility.

There is also a considerable angular distortion with respect to the
ideal tetrahedral angle. Not surprisingly, the largest angle
(123.9$^\circ$) occurs within the heptagonal rings (two other
different angles in the sevenfold rings are close by, at
120.6$^\circ$; cf.\ the angle of a regular heptagon of
128.57$^\circ$).  The minimum angle (92.6$^\circ$) occurs at the
vertex of the fivefold rings (where the rest of the angles vary
between 101.4$^\circ$ and 102.7$^\circ$).  In spite of these
distortions it is quite remarkable that the Si1 sites at the junction
of two adjacent sixfold rings have a tetrahedral environment of
neighbors very similar to that of diamond sites.  In fact, the sixfold
rings themselves are very similar to those present in ideal diamond,
with angles deviating slightly from the ideal tetrahedral angle (from
107.7$^\circ$ to 110.6$^\circ$) and a mean angle of 109.3$^\circ$.
The angular deviations with respect to diamond are also minimal along
the three different zigzag chains running perpendicular to the $xy$
plane (108.6$^\circ$, 109.2$^\circ$, and 110.1$^\circ$) where the bond
lengths are also very similar to diamond.  As in P4$_2$/ncm, the bond
angles remain very stable under compression with a maximum variation
of 1.8$^\circ$ from -5 GPa to 10 GPa, but in fact most angles vary
only by tenths of a degree.

\subsubsection{The P4$_1$2$_1$2 structure: A new chiral framework with
  tetragonal symmetry}

Whereas Pbam and P4$_2$/ncm share some common elements with the
diamond structure, P4$_1$2$_1$2 does not show any obvious similarity
with diamond, beyond being also a tetrahedral network. With three
different types of site, six different bond lengths and sixteen
different bond angles, such a network is quite unlike the regular
tetrahedral network found in diamond. Instead, P4$_1$2$_1$2 has an
obvious similarity to the chiral framework structure (CFS) of $P6_522$
symmetry previously discovered by Pickard and Needs.
\cite{Pickard_2010_CFS_structure,footnote_2}
 
\begin{figure}
 \includegraphics[width=0.4\textwidth]{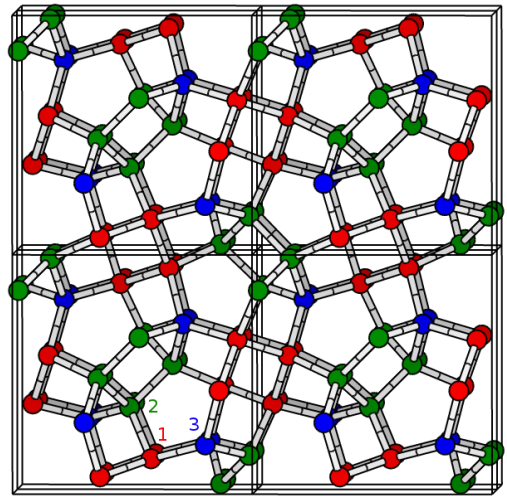}
\caption{\label{fig:structures_P4_12_12} (Color online) A slab of the
  P4$_1$2$_1$2 structure of Si at zero pressure, viewed along the
  direction of its fourfold rotation axis.  A modicum of perspective
  has been introduced to facilitate visualization of the connectivity
  among sites.  The three inequivalent sites are labelled and shown in
  different colours. }
\end{figure}

In Fig.\ \ref{fig:structures_P4_12_12} we show the P4$_1$2$_1$2
structure viewed from the direction of its main rotation axis (cf. 
the Supplemental Material\cite{Supplemental} and 
Fig.\ 1 of Ref.\ \onlinecite{Pickard_2010_CFS_structure} for a similar
plot of the CFS-P6$_5$22 structure).  The two-dimensional pattern of
this projection corresponds to the so called Cairo pentagonal tiling,
which is reported to appear in Islamic decoration.\cite{cairo_tiling}
Like the CFS-P6$_5$22 structure, the new P4$_1$2$_1$2 structure can be
viewed as made of interconnected spiral chains of bonds running both
clockwise and anti-clockwise along the $z$ axis, with some portions of
open space without clearly defined cavities (as, for example, in the
known clathrate structures).  In P4$_1$2$_1$2 there is one type of
threefold spiral, three types of fourfold spirals and one type of
fivefold spiral (whereas in CFS-P6$_5$22 there is one type each of
threefold, fourfold and sixfold spirals).  The presence of fivefold
spirals (Si1-Si2-Si2-Si1-Si3) in P4$_1$2$_1$2 is a rather unusual
feature, which is unique and characteristic of this structure among
others of group-14 elements previously reported or that we have found
in our searches.  There is also a large proportion of fivefold rings
(of which there are two types) as well as sixfold rings.  Angular
distortions from the ideal tetrahedral angle are considerable, with
the maximum angle (126.9$^\circ$) appearing along the fivefold spiral
chains (Si1-Si3-Si1) and the minimum angle (99.7$^\circ$) for
Si1-Si2-Si2.  Distances range from 2.35 to 2.39 \AA \ (cf. the value
of 2.37 \AA \ for the diamond phase).  The response of bonds to
compression appears to be similar to diamond with just one bond having
a somewhat larger compressibility.  There is a large variation in the
orientation of these bonds.

Both CFS-P6$_5$22 and CFS-P4$_1$2$_1$2 (which we could rename CFS-6
and CFS-5, respectively, on account of the order of the main spirals,
in each case) have rather similar energies, but the equilibrium volume
of CFS-P6$_5$22 is larger as it is based on sixfold instead of
fivefold spirals, which results in a more open structure, with larger
portions of empty space.  Though they have not appeared during our
structural searches, one may legitimately wonder whether other chiral
structures of the same kind but based upon sevenfold or even higher
order spirals would be energetically competitive and feasible as
stable phases.

\begin{table}
\centering
\caption{
  The equilibrium volume per atom, $V_0$, bulk modulus, $B_0$, pressure 
  derivative of the bulk modulus, $B_0^{\prime}$, and the difference in 
  energy from the corresponding diamond phase at zero pressure, 
  $\Delta E$, for various structural phases of carbon. 
% Experimental data (Exp.) are given where available.  
  These calculated results were obtained with 
  the \textsc{vasp} code and PBE density functional.
  (See also Supplemental Material for LDA and PBEsol results, as well as other phases.)
} 
\begin{ruledtabular}
\begin{tabular}{lcccc} 
{\bf Carbon}      & $V_0$ (\AA $^3$)  &  $B_0$ (GPa)   &   $B_0^{\prime}$   & $\Delta E$ (meV) \\
\hline
cd                & 5.708 & 435 & 3.51 & 0  \\
\hline
hd                & 5.722 & 433 & 3.60 & 25 \\
\hline
Pbam              & 5.873 & 416 & 3.60 & 69 \\
\hline
P4$_1$2$_1$2      & 5.897 & 413 & 3.61 & 132 \\
\hline
P4$_2$/ncm        & 5.971 & 403 & 3.64 & 108 \\
\hline
Cmca              & 5.874 & 413 & 3.72 & 128 \\
\hline 
Cmmm              & 5.870 & 413 & 3.67 & 129 \\
\hline
P6$_5$22          & 6.232 & 389 & 3.51 & 110 \\
\hline
I4$_1$/a          & 5.696 & 346 & 4.18 & 858 \\
\hline
bc8               & 5.616 & 386 & 3.84 & 697 \\
\hline
r8                & 5.658 & 363 & 3.98 & 819 \\
\hline
st12              & 5.650 & 395 & 3.74 & 886 \\
\hline
Ibam              & 5.771 & 372 & 3.64 & 956 \\
\hline
cintet            & 5.498 & 428 & 3.72 & 1115 \\
\hline
clat34            & 6.602 & 372 & 3.58 &  72 \\
\hline
clat46            & 6.546 & 368 & 3.61 & 106 \\
\end{tabular}
\end{ruledtabular}
\label{table:properties_carbon}
\end{table}

\begin{table}
\centering
\caption{
As Table \ref{table:properties_carbon}, but for silicon.
}
\begin{ruledtabular}
\begin{tabular}{lcccc} 
{\bf Silicon}     & $V_0$ (\AA $^3$)  &  $B_0$ (GPa)   &   $B_0^{\prime}$   & $\Delta E$ (meV) \\
\hline
cd                & 20.45 & 88.9  & 4.1 & 0   \\
\hline
hd                & 20.43 & 88.6  & 4.4 & 11 \\
\hline
Pbam              & 20.98 & 85.0 & 4.2 &  29 \\
\hline
P4$_1$2$_1$2      & 20.87 & 85.0 & 4.1 &  41 \\
\hline
P4$_2$/ncm        & 21.11  & 83.2 & 4.4 &  45   \\
\hline
Cmca              & 20.96 & 84.5 & 4.2 &  61    \\
\hline
Cmmm              & 20.95 & 84.8 & 4.2 &  61  \\
\hline
P6$_5$22          & 22.18 & 80.9 & 3.0 &  53 \\
\hline 
I4$_1$/a          & 18.17 & 77.1 & 4.4 & 165   \\
\hline
bc8               & 18.45 & 83.6  & 4.2 & 159 \\
\hline
r8                & 18.22 & 78.8  & 4.3 & 160 \\
\hline
st12              & 18.35 & 71.3 & 3.1 & 166 \\
\hline
Ibam              & 18.30 & 59.1 & 2.8 & 204   \\
\hline
cintet            & 18.12 & 96.0 & 4.4 & 203 \\ 
\hline
clat34            & 23.56 & 76.1 & 4.3  &  52 \\
\hline
clat46            & 23.27 & 75.8 & 4.2  &  67 \\
\end{tabular}
\end{ruledtabular}
\label{table:properties_silicon}
\end{table}

\begin{table}
\centering
\caption{
As Table \ref{table:properties_carbon}, but for germanium.
}
\begin{ruledtabular}
\begin{tabular}{lcccc} 
{\bf Germanium}   & $V_0$ (\AA $^3$)  &  $B_0$ (GPa)   &   $B_0^{\prime}$   & $\Delta E$ (meV) \\
\hline
cd                & 24.18 & 59.2  & 4.6 & 0 \\
\hline
hd                & 24.13 & 59.5  & 4.6 & 17 \\
\hline
Pbam              & 24.83 & 56.2  & 4.8 &  31   \\
\hline
P4$_1$2$_1$2      & 24.58 & 57.5  & 4.8 &  39   \\
\hline
P4$_2$/ncm        & 24.82 & 56.7  & 4.7 &  35   \\
\hline
Cmca              & 24.81 & 56.1  & 4.7 &  55  \\
\hline
Cmmm              & 24.81 & 56.3  & 4.6 &  54  \\
\hline
P6$_5$22          & 26.04 & 53.2  & 4.7 &  33  \\
\hline
I4$_1$/a          & 21.74 & 52.8  & 5.0 & 146   \\
\hline
bc8               & 22.22 & 56.5  & 4.4 & 137 \\
\hline
r8                & 21.83 & 53.6  & 4.9 & 142 \\  
\hline
st12              & 21.73 & 49.2  & 4.7 & 139  \\
\hline
Ibam              & 21.92 & 41.0  & 5.0 & 165  \\
\hline
cintet            & 21.94 & 61.7  & 4.7 & 173  \\
\hline
clat34            & 27.50 & 51.9  & 4.8 &  25  \\
\hline 
clat46            & 27.10 & 51.5  & 4.8 &  32  \\
\end{tabular}
\end{ruledtabular}
\label{table:properties_germanium}
\end{table}

\subsection{Band structures}
\label{results_bandstructures}

The band structures and electronic densities of states (EDoS) of the
Pbam, P4$_1$2$_1$2, and P4$_2$/ncm structures of Si at zero pressure
are shown in Fig.\ \ref{fig:Si_bandstructures}, and similar data for C
and Ge are provided in the Supplemental Material.\cite{Supplemental}
The calculated band gaps are expected to be underestimated, as is usual in DFT
calculations using semi-local density functionals such as PBE, but the
differences between the calculated band gaps are expected to be more
accurate, by which useful estimations can be obtained.

\begin{figure}[h!]
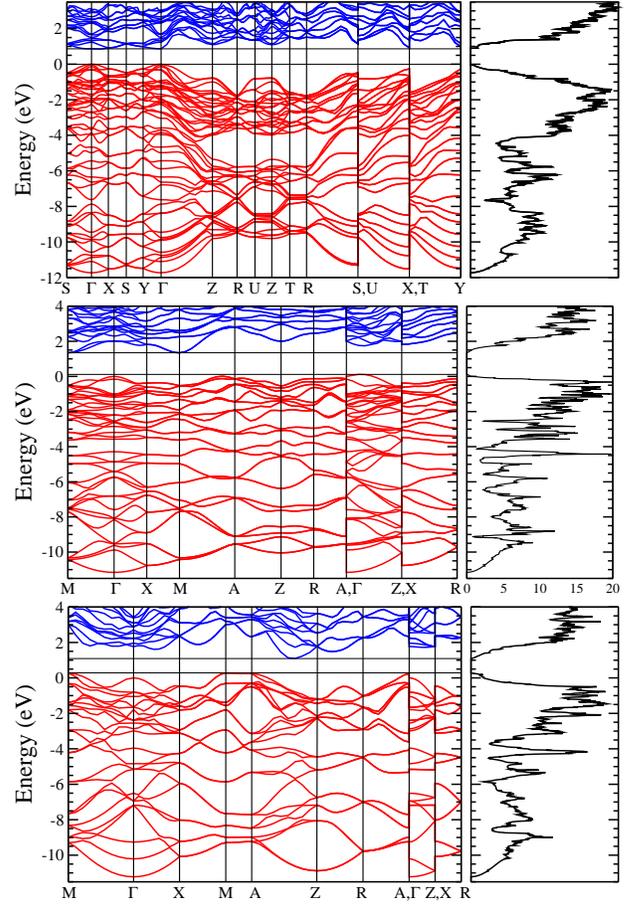

  \includegraphics[width=0.45\textwidth]{mujica_fig_8a.eps}
  \includegraphics[width=0.45\textwidth]{mujica_fig_8b.eps}
  \includegraphics[width=0.45\textwidth]{mujica_fig_8c.eps}
  \caption{\label{fig:Si_bandstructures} (Color online) Electronic
    band structures and densities of states of the Pbam, P4$_1$2$_1$2,
    and P4$_2$/ncm structures of Si at zero pressure.  The valence
    bands are shown in red and the conduction bands in blue, and the
    band gap regions are indicated by parallel black lines.  The zeros
    of energy are at the top of the valence band at the zone center
    $\Gamma$ point.  The panel on the right of each plot shows the
    total density of states.  }
\end{figure}

The C-Pbam, C-P4$_1$2$_1$2, and C-P4$_2$/ncm phases are insulators
with minimum band gaps calculated to be 
%in the range 3.70--4.70 eV,
4.57, 4.70, and 3.74 eV, respectively (cf. the calculated band gap 
of C-cd of 4.13 eV).
The occupied bandwidths are 20.8 eV (C-Pbam), 19.7 eV
(C-P4$_1$2$_1$2), and 19.9 eV (C-P4$_2$/ncm), 
%respectively, 
which are
somewhat smaller than the calculated bandwidth of the diamond phase,
21.4 eV (cf.\ the experimental value of of the bandwidth of C-cd,
23.0$\pm$0.2 eV, Ref.\ \cite{Jimenez_bandwidth_diamond_1997}).

The corresponding minimum band gaps for the Si structures are 0.86 eV
(Si-Pbam), 1.23 eV (Si-P4$_1$2$_1$2), and 0.80 eV (Si-P4$_2$/ncm),
which are larger than the gap of 0.63 eV calculated for the diamond
structure using PBE-DFT, which is in turn approximately half the
experimental value of 1.17 eV.  If we add to the band gaps of Si-Pbam,
Si-P4$_1$2$_1$2, and Si-P4$_2$/ncm a correction equal to the
difference between the theoretical and experimental band gaps of Si-cd
we obtain band gaps of about 1.4 eV (Si-Pbam), 1.77 eV
(Si-P4$_1$2$_1$2), and 1.34 eV (Si-P4$_2$/ncm).
%, respectively.  
These gaps are larger than in Si-cd, and are considerably closer to the values
around 1.5 eV that are optimal for photovoltaic applications.

In particular, the top of the valence band of Si-Pbam is located at the zone
center $\Gamma$, where the (corrected) direct band gap is $\sim$1.4 eV, and
therefore this phase could be of technological interest. The bottom of the
conduction band has a rather flat variation around $\Gamma$, along the
$\Gamma$-$X$ and $\Gamma$-$Y$ directions (with energy differences of the order of
0.01 eV).  Due to the small axial ratios $c/a$ and $b/a$ of the Pbam cell, both
the $X$ and $Y$ points are close to the zone center in Si-Pbam, see Fig.\
\ref{fig:Si_bandstructures}.  The very small dispersion of the conduction bands
around $\Gamma$ suggests that the joint density of states for electron
excitation is large.  In contrast, the band structures of both Si-P4$_1$2$_1$2
and Si-P4$_2$/ncm show distinct indirect band gaps, as Si-cd.

The Ge-Pbam phase has a calculated direct band gap at the zone center close to
zero, while Ge-P4$_1$2$_1$2 and Ge-P4$_2$/ncm have calculated indirect gaps of
about 1.00 eV and 0.30 eV, respectively.  Analogous calculations for Ge-cd give
a band gap of approximately zero (cf. the experimental value of 0.67 eV),
thus the corrected gaps for the tetrahedral polymorphs 
%Ge-P4$_1$2$_1$2 and Ge-P4$_2$/ncm 
are also expected to be larger than for the Ge-cd phase, as found in Si. 

A common feature of the band structures of the low-density polymorphs
of group 14 phases is a weakening of the $sp$ hybridization arising
from the substantial deviations of the bond angles from the perfect
tetrahedral angle of 109.5$^{\circ}$, and a small reduction in the
occupied valence bandwidth.  The EDoS of the Si-Pbam, Si-P4$_1$2$_1$2,
and Si-P4$_2$/ncm phases are not divided into separate $s$ and $p$
parts, as found in P6$_5$22, see Fig.\ 2 of Ref.\
\onlinecite{Pickard_2010_CFS_structure}, although the overlaps of the
$s$ and $p$ parts of the EDoS in the Si structures are not large.  The
EDoS of Ge-P4$_1$2$_1$2 is broken into a lower $s$ part and an upper
$p$ part, with a small inner gap clearly visible at about -3.7 eV indicating
significant weakening of the $sp$ hybridization. The band structure of
Ge-P4$_2$/ncm is almost broken into $s$ and $p$ parts, 
while for Ge-Pbam there is significant overlap in the energy scale between 
%$s$ and $p$ parts 
both parts, giving a gapless valence EDoS. 

\section{Vibrational properties and dynamical stability}
\label{sec_phonons}

The phonon dispersion relations of the Pbam, P4$_1$2$_1$2, and P4$_2$/ncm
phases of Si at zero pressure, calculated using a supercell method with small
atomic displacements,\cite{Alfe_phon} are plotted in Fig.\
\ref{fig:phonons_si}, while those of C and Ge are given in the Supplemental
Material.\cite{Supplemental} These plots show that the phases are all
dynamically stable at zero pressure.  The Pbam phases have distinct and
relatively well resolved upper-frequency phonon bands which have rather small
dispersion across the Brillouin zone, resulting in a characteristic
high-frequency peak in their respective density of states (PhDoS).  The phonon band
structures and PhDoS of the three phases in Si and Ge phases are quite similar,
differing mainly by a scaling arising from the different atomic massses, which
reflects the similarity in bonding.  For the three materials, the PhDoS of the
P4$_2$/ncm phase extends to somewhat lower frequencies than the Pbam phase.

\begin{figure}[h!]
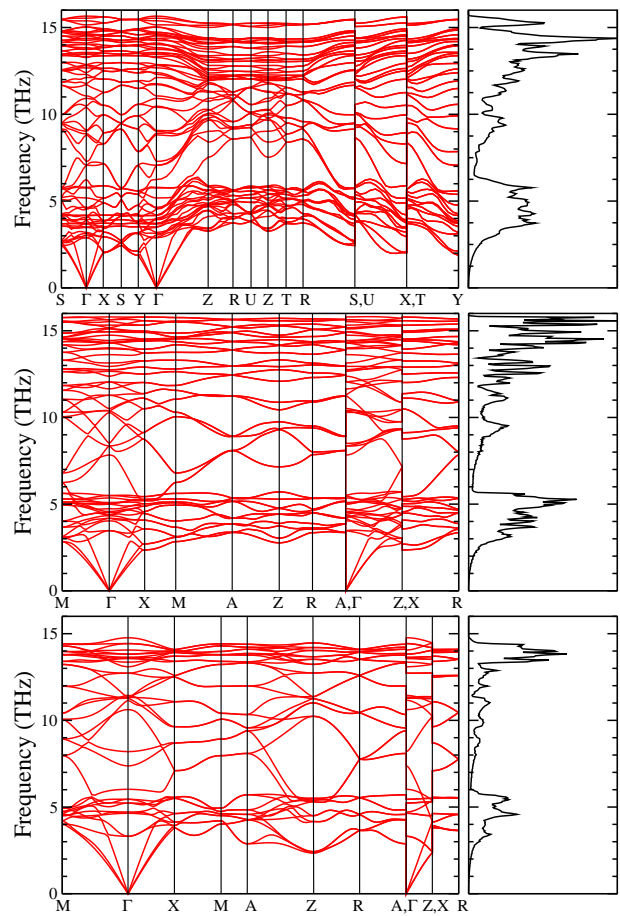

  \includegraphics[width=0.45\textwidth]{mujica_fig_9a.eps}
  \includegraphics[width=0.45\textwidth]{mujica_fig_9b.eps}
  \includegraphics[width=0.45\textwidth]{mujica_fig_9c.eps}
  \caption{\label{fig:phonons_si} (Color online) Calculated phonon dispersion
    relations of the Pbam, P4$_1$2$_1$2, and P4$_2$/ncm structures of
    Si at zero pressure.  The panel on the right of each plot shows
    the total phonon density of states.}
\end{figure}

\begin{figure}[ht!]
 \includegraphics[width=0.4\textwidth]{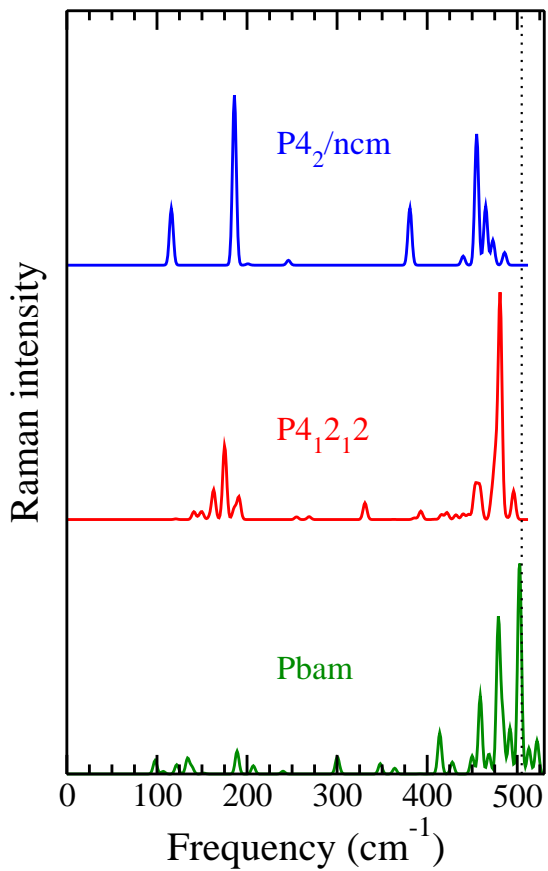}
 \caption{\label{fig:raman} (Color online) Calculated Raman spectra of the
   P4$_2$/ncm, P4$_1$2$_1$2, and Pbam structures of Si at zero pressure and 300 K,
   calculated within the PBE.  The vertical dotted line indicates the
   calculated frequency of the single intense Raman mode of the Si-cd phase.}
\end{figure}

We have also simulated the Raman intensities for the three phases at 300 K, which are
shown in Fig.\ \ref{fig:raman}.  Si-Pbam has strong high-frequency Raman
peaks in the range 450-510 cm$^{-1}$, and a number of rather weak Raman
modes below that frequency range.  Its strongest mode lies almost directly on
top of the calculated single Raman mode of Si-cd (505 cm$^{-1}$, cf. the experimental value of 
520 cm$^{-1}$, Ref.\ \onlinecite{Ruffell_2009}). This fact is
related to the structural similarity of Pbam and diamond discussed previously.
In contrast, the strongest Raman peak of Si-P4$_1$2$_1$2 lies at a frequency of
480 cm$^{-1}$, somewhat below the diamond mode.  Si-P4$_1$2$_1$2 shows
also significant Raman activity just below 200 cm$^{-1}$ and weak Raman
activity around 330 cm$^{-1}$, as can be seen in Fig.\ \ref{fig:raman}. 

These simulated spectra can be 
%complementarily 
used for the eventual
identification of the phases in Raman experiments, although insufficient
experimental resolution, broadened peaks, and mixture of phases with
overlapping peaks (all of them rather normal situations in nanoindentation or
conventional compression experiments) difficult the task of comparison with
experimental Raman results.
Also, some
caution has to be exerted as changing temperature can result in a significant 
change in the relative intensity of the calculated diffraction peaks.
%For example, our Raman frequencies for Si-P4$_2$/ncm are in good
%agreement with those reported by Zhao \textit{et al.},\cite{Zhao_2012} 
%but we find significant differences in the relative intensities of the
%modes, 
For example, we find significant differences in the relative intensities of the
modes reported for Si-P4$_2$/ncm by Zhao \textit{et al.},\cite{Zhao_2012}
probably due to the temperature used in simulating the Raman spectra. (Further
to this, Zhao {\it et al.} used the LDA while the results in Fig.\
\ref{fig:raman} were obtained using the PBE functional.)
%(We have performed Raman calculations using the LSDA functional and the
%corresponding relaxed structure and frequencies are in good agreement to those
%reported in Ref.\ \onlinecite{Zhao_2012}.)

Zhao \textit{et al.}\ \cite{Zhao_2012} also made the interesting
proposal that P4$_2$/ncm could explain the experimentally unknown 
structure of a Si-XIII phase, 
supported by a comparison of calculated and experimental Raman
vibrational data.
%and limited x-ray diffraction data.
\cite{Domnich_2002,Ge_2004,Domnich_2008,Ruffell_2009}  
Metastable Si-XIII is relevant to thermal processing of silicon wafers, which
induces large and potentially damaging stresses.
%Metastable Si-XIII 
It has been reported in several
sources\cite{Kailer_1997,Domnich_2002,Ge_2004,Domnich_2008,Ruffell_2009}
%Si-XIII 
%where it has been 
as being observed only in coexistence with other Si phases (bc8, r8, cd, and hd),
which makes it very difficult to determine its structure.  
The experimental situation concerning Si-XIII has been examined in detail in a
recent publication by Ruffell \textit{et al.}\ \cite{Ruffell_2009} who have carefully
reassessed previous experimental data on this
phase.\cite{Kailer_1997,Domnich_2002,Ge_2004,Domnich_2008} These authors
conclude the existence of three signature peaks which are characteristics of the Raman
spectrum of Si-XIII, at frequencies of 202, 333, and 478 cm$^{-1}$, which agree
quite well with those calculated by us for the P4$_1$2$_1$2 phase (cf. the data
in Fig.\ \ref{fig:raman} for Si-P4$_1$2$_1$2 with those in Fig. 6 of Ref.\
\onlinecite{Ruffell_2009} for the Si-XIII phase). In contrast, the Raman
spectrum of the P4$_2$/ncm structure, previously suggested for Si-XIII by Zhao
\textit{et al.},\ \cite{Zhao_2012} lacks the characteristic peak at 333
cm$^{-1}$, whereas their calculated peaks at $\sim$110 and $\sim$380 cm$^{-1}$
are absent in the experimental data for this phase. We suggest thus that
P4$_1$2$_1$2 is a more likely candidate for the structure of the Si-XIII phase.

\section{Summary and Conclusions}
\label{sec_summary}

Structure searches for group 14 elements using AIRSS have led to the
discovery of a number of low-energy structures, including low-density $sp^3$
bonded structures of $Pbam$ and $P4_12_12$ symmetries.  From the
structures that we have studied 
%with the PBE density functional 
in C,
Si, and Ge, only the cd and hd structures have significantly lower energies than
Pbam.  Another structure that readily appeared in our searches is the
P4$_2$/ncm (or T12) structure previously reported by Zhao {\it et
  al.}\cite{Zhao_2012} for which we provide a detailed interpretation
in terms of tilting and rebonding of diamond-like tetrahedra, that
complements and extends those authors' description. 
%We believe that this 
The P4$_2$/ncm structure is the simplest possible that can be built
by stacking slabs of tilted tetrahedra while preserving to a large
degree a highly regular fourfold coordination for the sites.
%(almost exactly regular for the Si1 sites and somewhat more
%distorted for the Si2 sites).
The Pbam, P4$_1$2$_1$2, and P4$_2$/ncm structures are dynamically stable
and have lower energies
than a number of metastable polymorphs of group 14 elements that have
been synthesized.  It is conceivable that one or more of these phases
could be formed as metastable phases in indentation or diamond anvil
cell experiments, perhaps in combination with heat treatment, 
and in fact P4$_2$/ncm has been proposed previously as the structure
of known yet experimentally unresolved phases of Si and Ge.\cite{Zhao_2012}
%of long known yet experimentally unresolved phases of Si and Ge.\cite{Zhao_2012}
We note that an equivalent P4$_2$/ncm structure 
has recently been suggested to play an important role in the
homogeneous crystallization of water.\cite{Russo_water}

Our PBE-DFT calculations show semiconducting behavior for the Pbam
and P4$_1$2$_1$2 
%and P4$_2$/ncm 
structures at zero pressure, except for
Ge-Pbam that is calculated to have a band gap close to zero at low
pressures.  The band gaps for a particular structure decrease from C
to Ge, as expected.  
The direct band gap of Si-Pbam at the zone center is estimated to have a value
of 1.4 eV (after correcting for the well-known underestimation of the PBE-DFT gap) 
so that it might be suitable for applications in photovoltaics.

We have found several other structures in our searches.
%, some of which have not, as far as we are aware, been reported before.  
These include a low-energy structure of $I4_1/a$ symmetry that is
denser than the cd structure in Si and has a very similar enthalpy to
that of the well-known metastable polymorphs r8/bc8, 
and which has been suggested to play a role in the decompression kinetics from the
high pressure $\beta$-Sn phase.\cite{wang_2013}
We also found a
low energy and low density structure of $Cmca$ symmetry which, in
carbon, is very close in energy to both the Pmmn structure of the so-called P
phase\cite{niu_2012} and the Cmmm structure or Cco-C8
phase,\cite{Zhao_2011_Cco-C8} with which it shares similar structural
features.  These phases have been proposed for superhard carbon
allotropes experimentally obtained after cold compression of carbon
nanotubes,\cite{wang_2004} and their closeness suggests that such experimental
allotropes could consist of a mixture of energetically and
structurally related forms.
%reexamination of previous results, thus.

Among the proposed structures, the new Pbam polymorph looks
particularly promising (low energy, dynamical stability and direct
band gap in Si).  Although the study of transformation
mechanisms between phases and possible synthesis routes is beyond the
scope of the present work, its structural features, based on sheared
diamond-like slabs, suggests that it might be synthesized by
controlled application of uniaxial stress to the diamond phases.
(Note added: When this paper was about to be submitted
we learnt of very recent findings for Pbam in C
by Baburin {\it et al.}\cite{Baburin_zeolite_to_sp3_carbon_2015}
Our results for C-Pbam agree with those in Ref.\ \onlinecite{Baburin_zeolite_to_sp3_carbon_2015}
while here we also provide results for Si and Ge.\cite{footnote_8})

P4$_1$2$_1$2, a spiral structure with tetragonal symmetry consisting of a
packing of pentagonal helices with a large proportion of fivefold rings, is
also very interesting and shows several unusual features. It belongs to the
same class of chiral frameworks as a previously reported P6$_5$22
structure\cite{Pickard_2010_CFS_structure} and its projection along the
four-fold axis yields the so-called Cairo two-dimensional pentagonal tiling. 
%which appears in Islamic decorations. 
%Quite surprisingly for us, 
%It is interesting that 
In a very recent publication\cite{zhang_2015} that appeared when this
paper was about to be submitted, a new two-dimensional carbon allotrope
with 24 atoms per cell and $P42_1m$ symmetry was proposed, with 
exactly the same description in terms of the Cairo tiling 
%structural motif 
as the three-dimensional P4$_1$2$_1$2 structure proposed here.
The appearance of the same unusual (yet beautiful) topological motif in both 
two- and three-dimensional metastable forms of carbon, obtained independently, 
%is quite interesting and begs for a closer examination of
is quite interesting and begs for a closer examination of
the role of pentagonal rings on the stability of novel and exotic carbon allotropes. 
The P4$_1$2$_1$2 structure is energetically quite competitive in Si and Ge, and in 
C it has a similar enthalpy to the structures currently proposed
for allotropes obtained from compressed nanotubes,\cite{niu_2012,wang_2004,Zhao_2011_Cco-C8}
yet with a very different topology.
We find that the P4$_1$2$_1$2 structure shows the best compatibility with the
available experimental Raman data for the unknown phase Si-XIII. 

We hope that these results will entice experimental searches for new
phases of these materials.

\begin{acknowledgments}
  AM acknowledges the financial support of the Ministerio de Educaci\'on,
  Cultura y Deporte (MECD, Spain) through its Programa de Movilidad de Recursos Humanos
  (Plan Nacional de I+D+i), grant PRX12/00335, and of project
  MAT2010-21270-C04-03. Access to the MALTA computer cluster (Universidad de
  Oviedo, Project CSD2007-00045) and the High Performance Computing Service of
  the University of Cambridge are gratefully acknowledged.  RJN and CJP were
  supported by the Engineering and Physical Sciences Research Council (EPSRC) of
  the UK.  We thank Keith Refson for useful discussions.
\end{acknowledgments}


\begin{thebibliography}{10}
\bibitem{footnote_7} We have also obtained several other
         diamond-like polytypes lying at energies similar to
         cubic and hexagonal diamond.
\bibitem{footnote_9} A second and denser state with the same $P4_2/ncm$ symmetry but lying 
         higher in enthalpy has also appeared in our survey.
\bibitem{footnote_4} 
         These are not planar rings: one half (blue) of each ring is in one 
%        (001) 
         plane and the other (red) in another plane a distance $c/2$ apart.
\end{thebibliography}
\end{document}